\documentclass[12pt]{article}
\usepackage{amssymb,amsmath}
\usepackage{graphicx}
\usepackage{epsfig}
\usepackage{epstopdf}
\usepackage{latexsym}
\usepackage{psfrag}
\usepackage{booktabs}
\usepackage{bbm}
\usepackage[toc]{appendix}
\usepackage{color}
\usepackage{datetime}
\usepackage[
      colorlinks=true,
      linkcolor=darkblue,  
      urlcolor=blue,    
      filecolor=blue,     
      citecolor=red,
linktocpage=true,
      pdfstartview=FitV,
      bookmarksopen=true    
      ]{hyperref}

\definecolor{darkblue}{rgb}{0.2, 0, 0.8}

\numberwithin{equation}{section}

\newcommand{\WZ}{\text{WZ}}

\newcommand{\UV}{\text{UV}}
\newcommand{\IR}{\text{IR}}
\newcommand{\m}{\mu}
\newcommand{\n}{\nu}
\renewcommand{\r}{\rho}
\newcommand{\s}{\sigma}
\renewcommand{\l}{\lambda}
\newcommand{\D}{\nabla}
\newcommand{\Da}{\,\Delta a\,}
\newcommand{\Dc}{\,\Delta c\,}
\newcommand{\mn}{{\mu\nu}}
\newcommand{\mnrs}{{\mu\nu\rho\sigma}}
\newcommand{\inv}{\text{inv}}
\newcommand{\pa}{\partial}
\newcommand{\sq}{\square}
\renewcommand{\t}{\tau}
\renewcommand{\b}{\beta}

\newcommand{\T}{\langle T_\m{}^\m\rangle}
\newcommand{\A}{\mathcal{A}}
\newcommand{\N}{\mathcal{N}}
\newcommand{\reef}[1]{(\ref{#1})}
\newcommand{\cn}{\mathcal{N}}

\newcommand{\cs}{\zeta}
\newcommand{\pA}{\varphi}
\newcommand{\pB}{\xi}

\newcommand{\compactsubsection}[1]{\vspace{0.1mm}\noindent {\bf #1}\\[1mm]}

\begin{document}  


\begin{titlepage}
\begin{flushright}
MCTP-13-43\\
\end{flushright}

\vspace*{2.3cm}

\begin{center}
{\Large\bf   Dilaton Effective Action with $\mathcal{N}=1$ Supersymmetry} \\

\vspace*{1.4cm}

{\bf Nikolay Bobev$^{1}$, Henriette Elvang$^{2}$, and Timothy M. Olson$^{2}$}
\bigskip
\bigskip

$^{1}$Perimeter Institute for Theoretical Physics \\
31 Caroline Street North, ON N2L 2Y5, Canada 
\bigskip

$^{2}$Randall Laboratory of Physics, Department of Physics,\\
University of Michigan, Ann Arbor, MI 48109, USA\\

\bigskip
nbobev@perimeterinstitute.ca, elvang@umich.edu, timolson@umich.edu  \\
\end{center}

\vspace*{5mm}

\begin{abstract}
We clarify the structure of the four-dimensional low-energy effective action that encodes the conformal and $U(1)$ R-symmetry anomalies in an $\cn=1$ supersymmetric field theory. The action depends on the dilaton, $\tau$, associated with broken conformal symmetry, and the Goldstone mode, $\beta$, of the broken $U(1)$ R-symmetry. We present the action for general curved spacetime and background gauge field up to and including all possible four-derivative terms. The result, constructed from basic principles, extends and clarifies the structure  found by Schwimmer and Theisen in \cite{Schwimmer:2010za} using superfield methods. We show that the Goldstone mode $\beta$ does not interfere with the  proof of the four-dimensional $a$-theorem based on $2 \to 2$ dilaton scattering. In fact, supersymmetry Ward identities ensure that a proof of the $a$-theorem can also be based on $2 \to 2$ Goldstone mode scattering when the low-energy theory preserves $\cn=1$ supersymmetry. We find that even without supersymmetry, a Goldstone mode for any broken global $U(1)$ symmetry cannot interfere with the proof of the four-dimensional $a$-theorem.

\end{abstract}

\end{titlepage}


{\small
\setlength\parskip{-0.5mm}
\tableofcontents
}

\section{Introduction and summary}

The dilaton-based proof \cite{Komargodski:2011vj,Komargodski:2011xv} of the four-dimensional $a$-theorem has provided new insights into the behavior of quantum field theories under renormalization group (RG) flows, for example in studies of conformal versus scale invariance \cite{Luty:2012ww,Dymarsky:2013pqa,Farnsworth:2013osa}. The arguments in \cite{Komargodski:2011vj,Komargodski:2011xv,Luty:2012ww} exploit that the structure of the effective action for the  dilaton --- introduced as a conformal compensator  or as the Goldstone boson for spontaneously broken conformal symmetry --- is determined by symmetries up to and including four-derivative terms. This is used to extract the change in the Euler central charge $\Delta a = a_\text{UV} - a_\text{IR}$ in an RG flow between UV and IR CFTs. The form of the dilaton action shows that the low-energy expansion of the scattering process of four dilatons is proportional to $\Delta a$ and a sum rule then allowed the authors of \cite{Komargodski:2011vj} to argue that $\Delta a >0$, thus proving the $a$-theorem. 

It is worth exploring if this argument can be affected by the presence of other massless modes in the low-energy theory, such as Goldstone bosons arising from the spontaneous breaking of other continuous global symmetries. This situation arises in $\N=1$ supersymmetric theories, because the stress tensor is in the same supermultiplet as the R-current, so the Goldstone boson $\beta$ for the broken $U(1)$ R-symmetry accompanies the dilaton $\tau$. In the low-energy effective action, there are couplings between $\tau$ and $\beta$, even in the flat-space limit, so one may wonder if this affects the proof of the $a$-theorem. 

Since the Goldstone boson $\beta$ is a pseudo-scalar (an axion), we are quickly relieved of our worries: its presence cannot change the scattering of four scalars (the dilatons) through single-axion exchanges, which would be the only option in the low-energy effective action. But precisely how this works is less trivial, since the ``naive" dilaton field $\tau$ is non-linearly coupled to the axion $\beta$, and to identify the physical modes one must disentangle the fields via a field redefinition. The result of course still holds true: the axion does not spoil the proof of the four-dimensional $a$-theorem presented in \cite{Komargodski:2011vj}.

In this note, we consider in detail the form of the bosonic terms in the $\N=1$ supersymmetric extension of the four-dimensional dilaton effective action in order to fully illuminate the above questions and to clarify results in the previous work \cite{Schwimmer:2010za}. Our focus  is four-dimensional $\mathcal{N}=1$ superconformal theories in which the conformal symmetry is broken by a relevant operator that preserves the $\N=1$ supersymmetry. We assume that the induced flow terminates in another $\N=1$ superconformal theory in the deep IR. The fields $\tau$ and $\beta$ form a complex scalar field which is the lowest component of a chiral Goldstone superfield $\Phi = (\tau + i\beta) + \dots$. We are interested in writing down the most general low-energy effective action for $\tau$ and $\beta$ in a general rigid four-dimensional curved space with background metric $g_{\m\n}$ and background $U(1)$ R-symmetry gauge potential $A_\m$. Such an action has been studied previously by Schwimmer and Theisen using a superspace approach \cite{Schwimmer:2010za}. One of our goals is to derive the action in component form from basic symmetry principles and use this to clarify the structure of the result presented in  \cite{Schwimmer:2010za}.

The fundamental ideas we use to determine the effective action $S[\tau,\beta]$ are diffeomorphism invariance and the following three properties:
\begin{enumerate}
\item Weyl variation $(\delta_\s g_{\mn}= 2\s g_\mn$ and $\delta_\s \t = \s)$  produces the trace anomaly, i.e.
\begin{align}
\delta_\sigma S = \int d^4x\, \sqrt{-g}\, \sigma\, \langle T_\mu{}^\mu \rangle \,.
\end{align}
The expectation value of the trace of the stress tensor, $\langle T_\mu{}^\mu \rangle$, is a functional of the background fields, namely the metric $g_\mn$ and the $U(1)_R$ gauge field $A_\mu$. It does not depend on $\tau$ or $\beta$. The full trace anomaly for an $\mathcal{N}=1$ SCFT with central charges $a$ and $c$ is\footnote{In Appendix \ref{app:anomaly} we discuss why no other terms involving the gauge field are allowed.}
\begin{align}
\langle T_\mu{}^\mu \rangle &= c W^2 - a E_4 + b' \square R - 6\,c\, (F_{\mu\nu})^2 \;.
\end{align}
The coefficient of $\sq R$ is non-physical as it can be removed by adding a local counterterm in the UV theory. Thus it is not an anomaly and we drop it henceforth. 
\item Gauge transformations $(\delta_\alpha A_\m = \D_\m \alpha$ and $\delta_\alpha \b = \alpha)$ generate the gauge anomaly:
\begin{equation}\label{gaugevary}
\delta_\alpha S 
= \int d^4x\, \sqrt{-g}\, \alpha\, 
\Big(2\,(5a-3c)\,F_{\mu\nu}\,\widetilde{F}^{\mu\nu} + (c-a)\,R_{\mu\nu\rho\sigma}\,\widetilde{R}^{\mu\nu\rho\sigma} \Big) 
\;,
\end{equation}
where the tilde denotes Hodge dualization with respect to the curved metric $g_{\mu\nu}$,
\begin{equation}
\widetilde{R}_\mnrs \equiv \frac{1}{2}\epsilon_{\mn\lambda\delta}R^{\lambda\delta}{}_{\r\s}\,, \qquad \widetilde{F}_\mn \equiv \frac{1}{2}\epsilon_{\mn\r\s}F^{\r\s}\;.
\end{equation}
The second line of \reef{gaugevary} gives the gauge anomaly\footnote{This is the 't Hooft anomaly for the global $U(1)_R$ symmetry present in any $\mathcal{N}=1$ SCFT. With slight abuse of notation we will refer to it as the gauge anomaly.} for the case of an $\mathcal{N}=1$ superconformal theory; it was derived in \cite{Anselmi:1997am} with slightly different normalization of $a$ and $c$ (see also \cite{Schwimmer:2010za,Intriligator:2003jj,Cassani:2013dba}).
\item The low-energy effective action must be invariant under $\N=1$ supersymmetry. Throughout this note we mostly ignore the fermionic degrees of freedom and focus entirely on the bosonic part of the action. 
 
\end{enumerate}

The first and second properties allow us to split the action into two parts $S = S_\WZ + S_\inv$ where Weyl  and gauge variations of $S_\WZ$ produce the trace  and gauge anomalies, respectively, while $S_\inv$ is gauge  and Weyl invariant. The general form of $S_\inv$ is a linear combination of all possible gauge  and Weyl invariant operators and the principles 1 and 2 above do not allow us to constrain the constant coefficients in this linear combination. However, the third property (supersymmetry) does fix certain relationships between the two parts of the action: some of the coefficients in $S_\inv$ are determined in terms of the central charges $a$ and $c$. This still leaves the possible freedom of having gauge  and Weyl invariant operators that are independently supersymmetric. We will show that no such operators contribute to the flat-space scattering process of four-particle dilaton and Goldstone modes at the four-derivative order. This means that such independently supersymmetric terms in the dilaton effective action (if they exist) cannot affect the proof of the $a$-theorem.

It is not easy to check whether a given four-derivative operator is supersymmetrizable. Thankfully the power of supersymmetry Ward identities allow us to test this question indirectly and to the extent we need it. As we show in Section \ref{sec:wards}, the supersymmetry Ward identities require that the scattering process of four dilatons is identical to the scattering process of the four associated R-symmetry Goldstone modes. This means that if an operator contributes only to  one of these processes, it cannot possibly be supersymmetrizable on its own. We use this to exclude contributions from Weyl  and gauge invariant operators that could otherwise affect the proof of the $a$-theorem in four-dimensional $\cn=1$ supersymmetric theories.\footnote{Very similar arguments were developed in \cite{Elvang:2010jv} to test supersymmetrization of candidate counterterms in $\N=8$ supergravity.}

Our work suggests several natural avenues for further exploration. First it will be interesting to analyze the effective actions for conformal field theories (not neccessarily supersymmetric) with larger continuous global symmetry groups. For superconformal theories with $\mathcal{N}=1$ supersymmetry and more than one Abelian global symmetry one may hope that such an effective action will offer a new perspective on the principle of $a$-maximization \cite{Intriligator:2003jj}. It will also be of great interest to construct the dilation effective action for four-dimensional SCFTs with extended supersymmetry, in particular for $\mathcal{N}=4$ SYM. In this context, one may be able to establish a more precise connection between the dilation effective action and the Dirac-Born-Infeld action for SCFTs with holographic duals. Finally, one can also study the supersymmetric dilation effective action for SCFTs in two and six dimensions.\footnote{There are no SCFTs in dimension greater than six and there are no conformal anomalies in odd dimensions. thus dimensions two, four and six exhaust all cases of interest.} The methods of this paper should extend readily to two-dimensional SCFTs with $(0,2)$ or $(2,2)$ supersymmetry since these theories have Abelian R-symmetry. The extension to six-dimensional $(1,0)$ or $(2,0)$ SCFTs may prove more subtle, although in the latter case holography should provide useful insights.

Before delving into the construction of the dilaton effective action, we start by deriving supersymmetry Ward identities for on-shell scattering amplitudes in Section \ref{sec:wards}. In Section \ref{sec:dea} we derive the most general form of the dilaton effective action for $\mathcal{N}=1$ SCFTs up to four-derivative terms. We compare this action to the results of Schwimmer-Theisen in Section \ref{sec:matchingST} to clarify the structure of their superspace-based result. In Section \ref{sec:amplitudes}, we show that the Ward identities from  Section \ref{sec:wards} confirm the supersymmetry of our result for the action in the flat-space limit.  The resulting dilaton-axion effective action gives an explicit verification that the dilaton-based proof is not affected by $\beta$. Furthermore, we show that supersymmetry is actually not needed to reach this conclusion: the Goldstone mode of any broken global $U(1)$ symmetry cannot spoil the proof of the $a$-theorem. Finally, we note that supersymmetry requires that the $2\to 2$ axion scattering amplitude must equal the $2 \to 2$ dilaton amplitude, and this allows for a proof of the $a$-theorem based on the axion scattering for $\mathcal{N}=1$ SCFTs. In Appendix \ref{app:anomaly}, we present a way to derive the conformal anomaly for four-dimensional CFTs from basic principles.

\section{Scattering constraints from supersymmetry}
\label{sec:wards}

Scattering amplitudes in supersymmetric theories  obey supersymmetry Ward identities \cite{Grisaru:1976vm,Grisaru:1977px}. We consider here an $\N=1$ chiral model with a complex scalar $\cs$ and its fermionic superpartner $\lambda$. In Section \ref{sec:amplitudes}, the chiral scalar will be related to the dilaton and $U(1)$ Goldstone modes. As a result of the supersymmetry transformations of the free fields, it can be shown \cite{Elvang:2013cua} that the supersymmetry generators $Q$ and $Q^\dagger$ act on the states as\footnote{We are abusing notation by using the same symbols to represent the fields and their corresponding creation and annihilation operators. Hopefully it is clear enough from context what we mean.}
\begin{align}
\begin{array}{rlcrl}
{[Q,\cs]} ~= & [p|\,\lambda \;, & \qquad & [Q^\dagger,\,\l] &=~ |p\rangle\,\cs \;,\\
{[Q,\,\l]} ~=& 0\;, && [Q^\dagger,\cs] &= ~0\;,\\
{[Q,\overline{\cs}]} ~=&0\;, & & [Q^\dagger,\,\overline{\l}] &=~0\;, \\
{[Q,\,\overline{\l}]} ~=& [p|\,\overline{\cs}\;, && [Q^\dagger,\overline{\cs}] &=~ |p\rangle\,\overline{\lambda}
\,,
\end{array}
\end{align}
where the (anti)commutators are graded Lie brackets. The two-component spinors $|p\rangle$ and $[p|$ represent components of the particle momentum in the spinor-helicity formalism.\footnote{See the reviews \cite{Elvang:2013cua,Dixon:2013uaa} for more details about the spinor-helicity formalism and supersymmetry Ward identities.} More precisely, the on-shell four-momentum $p_\m$ for a massless particle can be written in terms of a pair of two-component spinors $|p\rangle^{\dot{a}}$ and $[p|^b$ as
\begin{equation}
p_\mu\,(\overline{\s}^\m)^{\dot{a}b} = - |p\rangle^{\dot{a}}[p|^b  \;,
~~~~~\text{and}~~~~~
 p_\m\,(\s^\m)_{a\dot{b}}= - |p]_a\langle p|_{\dot{b}} \,.
\end{equation}
For two light-like four-vectors, $p^\mu$ and $q^\mu$, angle- and square-brackets are defined as
\begin{align}
  [pq] = [p|^a |q]_a \;,
  ~~~~~\text{and}~~~~~
  \langle pq \rangle = \langle p| _{\dot{a}} |q\rangle^{\dot{a}}\,.
\end{align}
These brackets are antisymmetric, $[ pq ] = - [ qp ]$ and $\langle pq \rangle = - \langle qp \rangle$, because spinor indices are raised and lowered with the two-dimensional Levi-Civita symbol.

Now assuming the vacuum is supersymmetric, i.e.$\,Q\boldsymbol{|0\rangle} = Q^\dagger \boldsymbol{|0\rangle}=0$, we can derive supersymmetry Ward identities for the amplitudes. For example (treating $\lambda$ and $\cs$ as creation operators),\footnote{We are not including explicit momentum labels, but assume that the first state in the list has momentum $p_1^\mu$, the next $p_2^\mu$ etc.}
\begin{align}
0=\boldsymbol{\langle0|}\,
\big[Q^\dagger,\l\,\cs\,\cs\,\cs \big]\,\boldsymbol{|0\rangle} 
= 
 \boldsymbol{\langle0|}\,
 \big[Q^\dagger,\l \big]\, \cs\,\cs\,\cs \,\boldsymbol{|0\rangle} 
= 
|p_1 \rangle \,
\boldsymbol{\langle0|}\,
\cs\, \cs\,\cs\,\cs \,\boldsymbol{|0\rangle} \;,
\end{align}
where we have used that $Q^\dagger$ annihilates $\cs$. This is simply the statement that at any loop-order, the on-shell four-scalar amplitude $\A_4(\cs\,\cs\,\cs\,\cs)$ must vanish (where now we mean the particles created by the field $\cs$). Similarly,  $\A_4(\overline{\cs}\,\overline{\cs}\,\overline{\cs}\,\overline{\cs})=0$.

The four-scalar amplitudes with three $\cs$ and one $\overline{\cs}$ also vanish. To see this, we write 
\begin{equation}
0 =
\boldsymbol{\langle 0|}\,\big[Q^\dagger,\overline{\cs}\,\l\,\cs\,\cs \big]\,\boldsymbol{|0\rangle} =
|p_1\rangle\,\boldsymbol{\langle 0|}\, \overline{\lambda}\,\l\,\cs\,\cs\,\boldsymbol{|0\rangle} +|p_2\rangle\,\boldsymbol{\langle 0|}\,\overline{\cs}\,\cs\,\cs\,\cs \big]\,\boldsymbol{|0\rangle}\,.
\label{SWI2}
\end{equation}
Now dot in $\langle p_1|$ and use the antisymmetry of the angle bracket to eliminate the first term on the right hand side in \reef{SWI2}. For generic momenta, this leads to the statement that $\A_4(\,\overline{\cs}\,\cs\,\cs\,\cs\, )= 0$.

A similar story applies to scalar amplitudes with three $\overline{\cs}$'s. Altogether, supersymmetry requires the following amplitudes to vanish:
\begin{equation}
\begin{split}
&\A_4(\cs\,\cs\,\cs\,\cs) 
~=~
\A_4(\overline{\cs}\,\overline{\cs}\,\overline{\cs}\,\overline{\cs}) ~=~ 0\,,
\\[1mm]
&\A_4(\overline{\cs}\,\cs\,\cs\,\cs)
~=~\A_4(\cs\,\overline{\cs}\,\cs\,\cs)
~=~ \ldots 
~=~ \A_4(\overline{\cs}\,\overline{\cs}\,\overline{\cs}\,\cs)
= 0\,.
\end{split}
\label{SWI3}
\end{equation}
The second line includes all four-point amplitudes with an odd number of $\cs$'s. Amplitudes with two $\cs$'s and two $\overline{\cs}$'s, such as $\A_4(\overline{\cs}\,\overline{\cs}\,\cs\,\cs)$, are permitted to be non-vanishing by supersymmetry. The reader may be puzzled: surely a supersymmetric Lagrangian can have interactions terms of the form $\cs^4 + \overline{\cs}^4$, so how can that be compatible with our claim above that for massless scalars $\A_4(\cs\,\cs\,\cs\,\cs) = 0$? To see this in an example, consider an $\N=1$ theory with a canonical kinetic term $\Phi^\dagger \Phi$ and a superpotential $W = f \Phi + \tfrac{1}{5}\Phi^5$. The scalar potential $V = |dW/d\cs|^2 = |f|^2 + f \cs^4 +\bar{f} \,\overline{\cs}^4 + \cs^4 \overline{\cs}^4$ has exactly the four-scalar interaction terms that our supersymmetry Ward identity argument appears to be incompatible with. However, the origin $\cs=\overline{\cs}=0$ is obviously not a supersymmetric vacuum, so the Ward identity --- which used $Q^\dagger \boldsymbol{|0\rangle} =0$ --- is not valid. If we expand around another vacuum, we generate mass-terms and we are only interested in the case of massless particles. This resolves the puzzle.

Now suppose we decompose the complex scalar field $\cs$ into its real and imaginary parts, $\cs = \pA + i \pB$ and denote the corresponding scalar, $\pA$, and pseudo-scalar, $\pB$, states by the same symbols.  Expanding the supersymmetry constraints \reef{SWI3} then leads to the following non-trivial constraints on the amplitudes:\footnote{The Ward identities also imply certain relationships between the four-point amplitudes containing only one $\pA$ or one $\pB$, 
e.g.~$\A_4(\pA\pA\pA\pB)=-\A_4(\pB\pB\pB\pA)$. These relations are independent from those in \eqref{wards1}--\eqref{wards2}. However, they are trivially satisfied for our application because any amplitude with an odd number of pseudo-scalars $\pB$  vanishes in a parity-invariant theory. }
\begin{align}
\A_4(\pA\,\pA\,\pA\,\pA) ~&=~ \A_4(\pB\,\pB\,\pB\,\pB)\,, \label{wards1}
\\
\A_4(\pA\,\pA\,\pA\,\pA) ~=~ \A_4(\pA\,\pA\,\pB\,\pB) \,&+\, \A_4(\pA\,\pB\,\pA\,\pB) \,+\, \A_4(\pA\,\pB\,\pB\,\pA)\,. \label{wards2}
\end{align}

These linear relations between amplitudes will be very valuable in the analysis of the ${\N=1}$ low-energy effective action for the dilaton. In this context, $\pA$ will be associated with the physical dilaton and $\pB$ with the R-symmetry Goldstone mode. Thus, without knowing any details of the form of the $\N=1$ supersymmetric dilaton effective action, we have already learned from the first identity \eqref{wards1} that the four-dilaton amplitude must  be equal to the four-axion amplitude. The second identity \eqref{wards2} is important for testing that the explicit action we derive in Section~\ref{sec:amplitudes} is supersymmetric. 

The identities in \eqref{wards1}--\eqref{wards2} can also be used to test if a given candidate Weyl  and gauge invariant operator is compatible with supersymmetry. If the on-shell four-point amplitudes resulting from the operator do not satisfy \eqref{wards1}--\eqref{wards2}, then the operator cannot be supersymmetrized.  On the other hand, if the resulting amplitudes are compatible with \eqref{wards1}--\eqref{wards2}, then the operator has a supersymmetric extension at the level of four fields (though not necessarily beyond that order).

\section{Dilaton effective action}
\label{sec:dea}

We turn now to the construction of an $\N=1$ supersymmetric effective action for the dilaton and axion fields $\t$ and $\b$ in the presence of a curved background metric $g_\mn$ and background gauge field $A_\m$. As noted in the Introduction, the dilaton effective action can be split into two parts
\begin{equation}
S = S_\WZ + S_\inv \;,
\label{Ssplit}
\end{equation}
depending on whether gauge and Weyl transformations act non-trivially.

\subsection{Wess-Zumino action}
\label{sec:WZaction}

The Wess-Zumino part of the action is defined such that its gauge variation produces the anomaly for the $U(1)_R$ symmetry and its Weyl variation results in the conformal anomaly. It can be obtained either by iteratively applying transformations and adding terms to cancel extra variations, or by integrating the anomalies directly \cite{Wess:1971yu}. The result is the four-dimensional Wess-Zumino action for the dilaton and axion:
\begin{equation}\label{SWZdef}
\begin{split}
S_\WZ =& \int d^4x\,\sqrt{-g}\,\bigg[ \Dc\,\tau\, W^2\,-\Da\tau\,E_4
- 6\,\Dc\,\tau\, F^2 \\
&\qquad\qquad~~~   + \beta\Big( 2\,(5\Da-3\Dc)F\widetilde{F} + (\Dc-\Da)R\widetilde{R}\Big) \\
&\qquad\qquad~~~-\Da\bigg(4\,\Big(R^{\mu\nu}-\frac{1}{2}R\,g^{\mu\nu}\Big)\nabla_\mu\tau\,\nabla_\nu\tau - 2\,(\nabla\tau)^2\Big(2\,\Box\tau - (\nabla\tau)^2\Big) \bigg) \bigg]\;.
\end{split}
\end{equation}
Here $F=dA$ is the flux for the background $U(1)_R$ gauge field. Under a Weyl transformation, the variation of $\tau$ on the first line produces the conformal anomaly, while the Weyl tensor and field strength are inert. However, $E_4$ is not inert, but the  Weyl variation of the third line cancels the contributions from $\tau\, \delta_\s (\sqrt{-g}E_4)$. The second line is Weyl invariant. Gauge transformations shift $\beta \to \beta + \alpha$, hence the second line in \reef{SWZdef}  produces the $U(1)_R$ anomaly. When the flux and the axion vanish, one recovers the WZ action for the dilaton \cite{Schwimmer:2010za,Komargodski:2011vj,Komargodski:2011xv}. The coefficients $\Da=a_\UV - a_\IR$ and $\Dc=c_\UV - c_\IR$ are the difference between the corresponding central charges of the UV and IR SCFTs, as required by the anomaly matching conditions \cite{Schwimmer:2010za,Komargodski:2011vj}. 

\subsection{Gauge and Weyl invariants}

Since $S_\WZ$ is determined by its variation, it is only specified up to terms whose gauge and Weyl variations vanish. We define $S_\inv$ to be the sum of all independent gauge and Weyl invariant combinations of $\tau$, $\beta$, $g_\mn$, and $A_\m$. To facilitate the analysis, we define a Weyl invariant metric $\hat{g}_\mn = e^{-2\tau}g_\mn$, so that any curvature terms computed in terms of $\hat{g}_\mn$ will be invariant. This procedure appeared in the analysis in \cite{Komargodski:2011vj} (see also \cite{Elvang:2012st,Elvang:2012yc,Baume:2013ika} for analogues in higher dimensions) where there were three possible four-derivative Weyl invariants with independent coefficients: $\sqrt{-\hat{g}}\hat{W}^2$, $\sqrt{-\hat{g}}\hat{R}^2$, and $\sqrt{-\hat{g}}\hat{E}_4$. (The Euler density $\hat{E}_4$ is total derivative in four dimensions so it can be dropped.) In the present context, the additional fields  can be used to construct invariants. Specifically, the combination $(A-\D\beta)_\m$ is both gauge and Weyl invariant. This combination also suggests that we should treat $A_\m$ on the same footing as a derivative in the low-energy effective action. With these building blocks we find the most general Ansatz for $S_\inv$ including terms with at most four derivatives:
\begin{equation}\label{Sinv}
\begin{split}
S_\inv = \int d^4x\,&\sqrt{-\hat{g}}\,\bigg[-\frac{f^2}{2}\,\bigg(\, \frac{\hat{R}}{6} + \hat{g}^{\mu\nu}\,(A-\nabla\beta)_\mu\,(A-\nabla\beta)_\nu \bigg) + \sum\limits_{i=1}^{9}\gamma_i\,W_i
 + \mathcal{O}(\D^6) \bigg] \;,
\end{split}
\end{equation}
where we have dropped total derivatives such as $\sqrt{-\hat{g}}\hat{E}_4$. The hatted two-derivative gauge-Weyl invariants produce the kinetic terms for the scalars when expanded in terms of the unhatted metric and the dilaton. The real constants $\gamma_1,\ldots,\gamma_9$ are arbitrary coefficients of the independent four-derivative gauge and Weyl invariant terms, $\sqrt{-\hat{g}}W_i$, defined by
\begin{align}\label{Weylinv}
\begin{array}{lcl}
W_1\equiv\hat{W}^2\;, & & W_2 \equiv\hat{R}^2\;, \\
W_3\equiv(A-\nabla\beta )_\mu\,\hat{\nabla}^\mu\hat{R}\;, & & W_4\equiv\Big(\hat{\nabla}^\mu(A-\nabla\beta)_\mu\Big)^2\;, \\
W_5 \equiv \hat{g}^{\mu\nu}\,(A-\nabla\beta)_\mu\, \hat{\square}\,  (A-\nabla\beta)_\nu\;, & & W_6 \equiv \hat{R}^{\mu\nu}\, (A-\nabla\beta)_\mu\,(A-\nabla\beta)_\nu\;, \\
W_7 \equiv \hat{R}\,\hat{g}^{\mu\nu}\, (A-\nabla\beta)_\mu\,(A-\nabla\beta)_\nu\;, &\hspace{1.5cm} & W_8 \equiv \Big(\hat{g}^{\mu\nu}\,(A-\nabla\beta)_\mu\,(A-\nabla\beta)_\nu\Big)^2\;,\\
\multicolumn{3}{l}{W_9 \equiv \hat{g}^{\mu\nu}\,(A-\nabla\beta)_\mu\,(A-\nabla\beta)_\nu\,\hat{\nabla}^\lambda(A-\nabla\beta)_\lambda\;.}\\
\end{array}
\end{align}
All other invariants can be written as linear combination of the $W_i$ and total derivatives, e.g. the Bianchi identity implies ${\hat{R}^{\mn}\, \hat{\D}_\m(A-\D\b )_\n = \hat{\D}_\m \Big(\hat{R}^\mn\,(A-\D\b )_\n \Big) - \frac{1}{2}W_3}$.

This is the most general possible action written in terms of natural gauge and Weyl invariant objects constructed from the basic fields. So far, we have not imposed any supersymmetry on the Weyl+gauge invariant action $S_\text{inv}$. 
As we will see in the following sections, the constraints implied by $\N=1$ supersymmetry and the consequences for the $a$-theorem are easily expressed and understood in terms of the $W_i$ and their coefficients.

\section{Matching to superspace calculation}
\label{sec:matchingST}

The bosonic terms in the $\cn=1$ supersymmetric version of the Wess-Zumino action were derived earlier by Schwimmer and Theisen \cite{Schwimmer:2010za}. They started with the Weyl anomaly in superspace and integrated it directly using the Wess-Zumino method \cite{Wess:1971yu}. This gives a superspace form of the Wess-Zumino action which was then expanded in component fields; the result is given in equation (3.23) of \cite{Schwimmer:2010za}. In that expression, it is easy to pick out the terms that match $S_\WZ$ in \reef{SWZdef}. The two-derivative terms in \reef{Sinv} are also easily recognized. However,  it is not \emph{a priori} clear how to interpret the rest of the 4-derivative terms in (3.23) of \cite{Schwimmer:2010za}. Indeed, at first sight it may seem almost miraculous that these additional terms would not contribute to the anomaly under a gauge/Weyl transformation. 

The correct interpretation of the rest of the terms in (3.23) of \cite{Schwimmer:2010za} is that they are a combination of gauge and Weyl invariants required for the  supersymmetric completion of $S_\WZ$ in \reef{SWZdef}. Thus, the extra terms in (3.23) of \cite{Schwimmer:2010za} are a particular linear combination of the operators $W_i$ from \reef{Weylinv}: there is a unique choice of $\gamma_i$ in $S_\inv$ \eqref{Sinv} such that our action $S=S_\WZ+S_\inv$ agrees with (3.23) in \cite{Schwimmer:2010za}.\footnote{Our sign conventions differ from those of \cite{Schwimmer:2010za}. We use the curvature convention $[\D_\m,\D_\n]\,V^\r=R_{\mn}{}^\r{}_\s\,V^\s$. All equations shown here can be translated into the conventions of \cite{Schwimmer:2010za} by flipping the signs of the curvature tensors. We use a different normalization for $f$ and $A_\m$, namely $f^2_{\text{here}} = 2f^2_{\text{ST}}$ and $A_{\text{here}} = \frac{2}{3}A_{\text{ST}}$. Also, our result \reef{S0} fixes minor typos in \cite{Schwimmer:2010za}.} This choice is to set 
\begin{equation}\label{gamma678}
\gamma_6 \,=\, -6\,\gamma_7 \,=\, 2\,\gamma_8 \,=\, -4\Da 
\end{equation}
and drop the other $W_i$'s. This yields the following action:
\begin{equation}\label{Sst}
\begin{split}
&S_0 = \int d^4x\,\bigg\{-f^2\,\sqrt{-\hat{g}}\,\bigg[\, \frac{1}{12}\hat{R} +\frac{1}{2}\Big(\hat{g}^{\mu\nu}\,(A-\nabla\beta)_\mu\,(A-\nabla\beta)_\nu\Big) \bigg]\\
&\qquad~~ +\sqrt{-{g}}\,\bigg[ \Dc\,\tau\, W^2\,-\Da\tau\,E_4
- 6\,\Dc\,\tau\, F^2 \\
&\hspace{4cm}   + \beta\Big( 2\,(5\Da-3\Dc)F\widetilde{F} + (\Dc-\Da)R\widetilde{R}\Big) \\
&\hspace{4cm} -\Da\bigg(4\,\Big(R^{\mu\nu}-\frac{1}{2}R\,g^{\mu\nu}\Big)\nabla_\mu\tau\,\nabla_\nu\tau - 2\,(\nabla\tau)^2\Big(2\,\Box\tau - (\nabla\tau)^2\Big) \bigg) \bigg]\\
&\qquad~~ - 4\Da\sqrt{-\hat{g}}\,\bigg[ \Big(\hat{R}^\mn - \frac{1}{6}\hat{R}\,\hat{g}^\mn \Big) (A-\nabla\beta)_\mu\,(A-\nabla\beta)_\nu\\
&\hspace{4cm} + \frac{1}{2}\,\Big(\hat{g}^{\mu\nu}\,(A-\nabla\beta)_\mu\,(A-\nabla\beta)_\nu\Big)^2 \bigg]
 + \mathcal{O}(\D^6)
\bigg\}\;.
\end{split}
\end{equation}
The first line contains the kinetic terms. The second through fourth lines are the WZ action, \eqref{SWZdef}, whose Weyl and gauge variations respectively produce the conformal and $U(1)_{R}$ anomaly. The last two lines are gauge and Weyl invariant and can be viewed as the supersymmetric completion of the Wess-Zumino action.

Although the other $\gamma_i$ and $W_i$ do not appear in \eqref{Sst}, this should not be interpreted as setting them equal to zero. Rather, the remaining $\gamma_i$ do not contribute to \eqref{Sst} because the superspace calculation in \cite{Schwimmer:2010za} derived only the terms related to the anomaly in a general $\N=1$ theory. At present, the rest of the $\gamma_i$ are not fixed. We will see later that the supersymmetry Ward identities imply additional constraints.

One can now expand \eqref{Sst} to facilitate comparison with equation (3.23) in \cite{Schwimmer:2010za}:
\begin{equation}\label{S0}
\begin{split}
S_0 &= -f^2\int d^4x\,\sqrt{-g}\,e^{-2\tau}\Big(\frac{1}{2}(\D\tau)^2 + \frac{1}{12}R + \frac{1}{2}\big(\D\beta - A \big)^2 \Big)\\
&\quad + \int d^4x\,\sqrt{-g}\,\Big[\Dc\tau\,W^2 - \Da\tau\,E_4 -  6\Dc\tau\,(F_\mn)^2\\
&\hspace{4cm} +\beta\,\Big(2\,(5\Da-3\Dc)\,F^\mn\,\widetilde{F}_\mn + (\Dc-\Da)\, R^\mnrs\,\widetilde{R}_\mnrs \Big)\Big]\\
&\quad + 8\Da \int d^4x\,\sqrt{-g}\,\bigg(\Big[R^{\mu\nu}A_\nu -\frac{1}{6}R\,A^\mu + A^2\,A^\mu \Big]\,\nabla_\mu\beta - A^\mu\, A^\nu\, \nabla_\mu\nabla_\nu\tau \bigg) \\
&\quad  +\,2 \Da \int d^4x\,\sqrt{-g}\, \bigg\{\bigg[ \Big(R+2\, A^2\Big)g^{\mu\nu} - 2\,\Big(R^{\mu\nu} + 2\,A^\mu\,A^\nu\Big) \bigg] \nabla_\mu\tau \,\nabla_\nu\tau \\
&\qquad~ + \,\bigg[ \Big(\frac{1}{3}R - 2\,A^2 \Big) g^{\mu\nu} - 2\,\Big(R^{\mu\nu} + 2\, A^\mu\,A^\nu\Big) \bigg] \nabla_\mu\beta\, \nabla_\nu\beta + 8\,A^\nu \nabla^\mu \beta \,\nabla_\nu\nabla_\mu  \tau 
\bigg\}
 + \ldots \;.
\end{split}
\end{equation}
Here the dots denote terms with either no $\beta$'s and $\tau$'s, or more than two of them. Higher-derivative terms are also suppressed. 

The comparison between our dilaton effective action and the result in \cite{Schwimmer:2010za} uniquely selects the three gauge-Weyl invariants $W_6$, $W_7$, and $W_8$ and fixes their coefficients as in  \reef{gamma678}. If there are any other gauge-Weyl invariants in the low-energy dilaton-axion effective action, then their linear combination must be independently supersymmetrizable. We analyze this in the next section.

\section{Dilaton and axion scattering in flat space}
\label{sec:amplitudes}

For the purposes of testing supersymmetry and investigating the $a$-theorem, we now take the theory on a flat background with vanishing gauge field. Then $\tau$ and $\beta$ will be the only fields involved. For the moment, we continue to ignore the other $W_i$ that did not contribute to \eqref{Sst}. We will explain later why this is justified. The action \eqref{Sst} encodes the familiar dilaton interactions, as well as new couplings to the axion $\beta$. These new interactions are present even in the flat-space limit with no background gauge field. Up to total derivatives, we find
\begin{equation}\label{flatST}
\begin{split}
S_0 = \int d^4x\bigg\{&-\frac{f^2}{2}\,e^{-2\tau}\Big[ (\partial\tau)^2 + (\partial\beta)^2 \Big]
+ 2\Da \Big[ 2\,\Box\tau\big((\partial\tau)^2- (\pa\beta)^2 \big)+4\, \Box\beta\,(\partial\tau\cdot\partial\beta)  \\
&\quad -4\,(\partial\tau\cdot\partial\beta)^2 -\left((\partial\tau)^2 - (\partial\beta)^2\right)^2 \Big]+\mathcal{O}(\pa^6)\bigg\}\;.
\end{split}
\end{equation}
The fields $\tau$ and $\beta$ are coupled already at the two-derivative level through $e^{-2\tau}(\partial\beta)^2$, so the equations of motion mix $\tau$ and $\beta$: 
\begin{equation}
\square \tau = (\partial\tau)^2 - (\partial\beta)^2\;, \qquad\text{and}\qquad
\square \beta = 2(\partial\tau \cdot \partial\beta) \;.
\end{equation}
%

\subsection{Field redefinition}

To facilitate the calculation of scattering amplitudes, we make a field redefinition to decouple the kinetic terms. This is easiest when we identify the complex scalar field $Z$ that produces the kinetic terms
\begin{equation}
Z\equiv e^{-(\tau+i\,\b)} ~~\Rightarrow ~~ |\pa Z|^2 = e^{-2\tau}\Big((\partial\tau)^2 + (\partial\beta)^2\Big)\;.
\label{redef}
\end{equation}
The action (\ref{flatST}) can be rewritten in terms of $Z$ and its complex conjugate $\overline{Z}$ and takes a very simple form
\begin{align}
  S_0 = \int d^4x \,\bigg\{
       -\frac{f^2}{2}  \Big| \pa Z \Big|^2 + 2\Delta a 
       \bigg[
          -  \bigg(\frac{\pa Z}{Z}\bigg)^2 \frac{\Box \overline{Z}}{\overline{Z}}
          - \bigg(\frac{\pa \overline{Z}}{\overline{Z}}\bigg)^2 \frac{\Box {Z}}{{Z}}
          +\bigg|\frac{\pa Z}{Z}\,\bigg|^4 \,
       \bigg]
       +\mathcal{O}(\pa^6)
  \bigg\}\,.\label{STactZ}
\end{align}
Note that when the Goldstone mode $\beta$ vanishes we have a real scalar $Z \to e^{-\tau} \equiv \Omega$ and the action (\ref{STactZ}) reduces to the familiar form for the dilaton effective action in the flat space limit (see, for example, equation (2.8) in \cite{Luty:2012ww}). 

The field $Z$ is the compensator we introduce to restore the broken symmetries. We can expand about its constant vev\footnote{Note that one can always choose the vev of $Z$ to be real using the global $U(1)$ symmetry in the action \eqref{STactZ}.} $f$ with the fluctuating field $\cs$,
\begin{equation}
Z=1-\frac{\cs}{f}\, ,\qquad\qquad \cs = \pA + i\,\pB\;,
\end{equation}
where $\pA$ and $\pB$ are real scalar fields. Plugging this into the action \eqref{STactZ} and expanding up to fourth order in the fields, we find
\begin{equation}\label{phiST}
\begin{split}
S_0\to \int d^4x \,&\bigg\{ -\frac{1}{2}\bigg((\partial\pA)^2 + (\partial\pB)^2 \bigg) + \frac{4\Da}{f^3}\bigg( \sq\pA\, \Big((\pa\pA)^2 - (\pa\pB)^2 \Big)+ 2\,\sq\pB\,(\pa\pA\cdot \pa\pB) \bigg)\\
&\quad+ \frac{2\,\Delta a}{f^4}\bigg[2\, \sq\pA\,\Big(3\,\pA\,\Big((\pa\pA)^2-(\pa\pB)^2 \Big) -2\,\pB\,(\pa\pA\cdot\pa\pB) \Big) \bigg)\\
&\hspace{1.6cm} + 2\,\sq\pB\,\Big(\pB\,\Big((\pa\pA)^2-(\pa\pB)^2 \Big) +6\,\pA\,(\pa\pA\cdot\pa\pB) \Big) \bigg)\\
&\hspace{1.6cm} + \Big((\pa\pA)^2 - (\pa\pB)^2 \Big)^2 + 4\,(\pa\pA\cdot\pa\pB)^2\bigg] + \mathcal{O}(\pa^6)\bigg\}\;.
\end{split}
\end{equation}
This parameterization decouples the equations of motion into those of free massless scalars
\begin{equation}\label{EOMphixi}
\sq\pA = 0\,,\qquad\qquad \sq\pB = 0 \;.
\end{equation}
As an effective action with a derivative expansion, we only include the two-derivative quadratic terms in the equations of motion. All other terms in the action involve three or more fields and give rise to interaction terms in the quantized theory. In \eqref{phiST}, all such interactions involve at least four derivatives, so the amplitudes have no local contributions from pole diagrams until at least $\mathcal{O}(p^6)$.

\subsection{Amplitudes}

We are interested in the four-point amplitudes. From the action \eqref{phiST}, we see that the low-energy expansion starts at $\mathcal{O}(p^4)$. The equations of motion \eqref{EOMphixi} make it easy to read off the amplitudes from the contact terms in the last line of \eqref{phiST}, which yield at $\mathcal{O}(p^4)$:
\begin{equation}\label{STamps}
\begin{split}
\A_4(\pA\,\pA\,\pA\,\pA) &= \frac{4\Delta a}{f^4}(s^2+t^2+u^2)\;,\\
\A_4(\pB\,\pB\,\pB\,\pB) &= \frac{4\Delta a}{f^4}(s^2+t^2+u^2)\;,\\
\A_4(\pA\,\pA\,\pB\,\pB) &= \frac{4\Delta a}{f^4}(-s^2+t^2+u^2)\;,\\
\A_4(\pA\,\pB\,\pA\,\pB) &= \frac{4\Delta a}{f^4}(s^2-t^2+u^2)\;,\\
\A_4(\pA\,\pB\,\pB\,\pA) &= \frac{4\Delta a}{f^4}(s^2+t^2-u^2)\;.
\end{split}
\end{equation}
We can now use these results to check if the action \eqref{STactZ} is compatible with supersymmetry. 
Combining the corresponding results from \eqref{STamps}, we see that indeed the  constraints \eqref{wards1}--\eqref{wards2} from the supersymmetry Ward identities are obeyed. 

All three Weyl invariants, $W_{6,7,8}$, contributed to the amplitudes \eqref{STamps} in a non-trivial way that ensures that the supersymmetry Ward identities are satisfied. Hence, this tests the supersymmetry of \eqref{Sst}. The combination of Weyl invariants $W_i$ in \eqref{Sst}-\reef{S0} was fixed via comparison with the superspace form given by Schwimmer and Theisen \cite{Schwimmer:2010za}. The match was obtained by comparing the last three lines of \eqref{S0} with the corresponding expressions in \cite{Schwimmer:2010za}. Note that all the terms used explicitly in the match vanish in the flat-space limit with the background gauge potential turned off. However, as we have seen, $W_{6,7,8}$ also have flat-space contributions, so supersymmetry could also be tested via the Ward identities. 
Thus, in that limit, we have tested that our completion of the Schwimmer-Theisen terms does obey the supersymmetry constraints.

\subsection{Supersymmetry and the other Weyl invariants}

So far we have considered only the part of the action that matched the superspace derivation of the Wess-Zumino action, fixing the values of $\gamma_6$, $\gamma_7$, and $\gamma_8$. The full dilaton effective action may have contributions from the other invariants $W_i$ as well. This is important because their flat-space limits could include additional dilaton and axion scattering beyond what we have considered so far, with potentially dangerous consequences for the $a$-theorem.

With that in mind, let us return to the list of gauge-Weyl invariants \eqref{Weylinv} and evaluate them in the flat background. Applying the equations of motion \eqref{EOMphixi}, we find:
\begin{align}\label{Wiflat}
\begin{array}{l l}
W_1\to 0 \;,
&
W_2\to \frac{36}{f^4}(\pa\pB)^4\;,
\\
W_3\to 0\;,
&
W_4\to 0\;,
\\
W_5\to -\frac{2}{f^4}\Big( (\pa\pB)^4 + (\pa\pA\cdot\pa\pB)^2 \Big) \;,
\hspace{0.5cm}&
\boldsymbol{W_6}\to \boldsymbol{-\frac{2}{f^4}\Big( (\pa\pB)^4 + (\pa\pA\cdot\pa\pB)^2 \Big) }\;,
\\
\boldsymbol{W_7}\to \boldsymbol{-\frac{6}{f^4}(\pa\pB)^4 }\;,
&
\boldsymbol{W_8}\to \boldsymbol{\frac{1}{f^4}(\pa\pB)^4 }\;,
\\
W_9\to 0\;,
\end{array}
\end{align}
where the three expressions in boldface are those already included in \eqref{Sst}.

The first key feature to notice is that none of the invariants contain a $(\pa \pA)^4$ interaction.  Hence the four-scalar amplitude, $\A_4(\pA\,\pA\,\pA\,\pA)$ in \eqref{STamps}, receives contributions only from the dilaton part of the Wess-Zumino action. It is completely blind to the presence of the axion. Thus it is not surprising that the resulting amplitude in \eqref{STamps} matches exactly the one found in \cite{Komargodski:2011vj}. Moreover, this implies that the proof of the $a$-theorem using the four-dilaton amplitude is unaffected by the presence of the axion.

The second key feature is that any gauge+Weyl+supersymmetry invariant four-derivative term has to be a linear combination of the $W_i$'s, say $\mathcal{W}=\sum_{i=1}^9 b_i W_i$. Since \reef{Wiflat} tells us that the four-dilaton amplitude has zero contribution from $\mathcal{W}$, the supersymmetry Ward identity \reef{wards2} requires $b_5 + b_6 =0$, and consequently \reef{wards1} enforces 
$b_2 - 6 b_7+ 36 b_8 = 0$. There are no constraints on the other $b_i$'s from four-particle supersymmetry Ward identities. In conclusion, any gauge+Weyl+supersymmetry invariant four-derivative operator (if it exists) does not contribute at all to the four-particle scattering processes, so from that point of view we can completely neglect it. 

Using general principles, we have shown that --- up to four-derivative terms --- the dilaton-axion effective action for $\mathcal{N}=1$ SCFTs  takes the form $S=S_{\text{WZ}}+S_{\text{inv}}$, with $S_{\text{WZ}}$ and $S_{\text{inv}}$ given by \eqref{SWZdef} and \eqref{Sinv} respectively. The results of \cite{Schwimmer:2010za} fix the coefficients $\gamma_i$ as in \eqref{gamma678} to complete the Wess-Zumino action to an $\cn=1$ supersymmetric form. The supersymmetry Ward identities can be applied in the flat-space limit to see that no supersymmetric linear combination of the $W_i$'s contribute to any four-particle process. However, we cannot eliminate the possibility of such supersymmetric combinations; we can only say that in the flat-space limit their four-field terms must be proportional to total derivatives and the EOM. It would  be curious to know if such fully supersymmetric operators do exists, although we have established that for the proof of the $a$-theorem in four dimensions they do not matter.

We have demonstrated that the four-point axion scattering amplitude is given by the second line in \eqref{STamps}. One can now use the same positivity arguments as in \cite{Komargodski:2011vj,Komargodski:2011xv} to show that for $\mathcal{N}=1$ SCFTs $\Delta a=a_{\text{UV}}-a_{\text{IR}} >0$. This can be regarded as an alternative route to the $a$-theorem for four-dimensional SCFTs with $\mathcal{N}=1$ supersymmetry. 

\subsection{No supersymmetry}
Suppose we do not assume $\cn=1$ supersymmetry. Then the coefficients in the gauge anomaly \reef{gaugevary} are no longer fixed in terms of the trace anomalies $a$ and $c$. This affects only the second line of the WZ action \reef{SWZdef}, now with $\beta$ interpreted as the Goldstone mode of \emph{some} broken $U(1)$ symmetry. Nothing else changes in the WZ action.  The general form of the Weyl  and gauge invariant action \reef{Sinv} is unchanged in the flat-space limit with $A_\mu=0$. (The relative normalization between $A_\mu$ and $\beta$ may change, but we do not have to worry about this when $A_\mu=0$.) Of course, there is no supersymmetry or other principle to fix the coefficients $\gamma_i$. However, that is not important for the Komargodski-Schwimmer proof of the $a$-theorem because \reef{Wiflat} shows that none of the Weyl+gauge invariants $W_i$ affect the $2 \to 2$ scattering amplitude of the physical dilaton  at order $p^4$. Hence we conclude that even in the absence of supersymmetry the proof of the $a$-theorem is unaffected by the presence of Goldstone bosons for Abelian global symmetries. 

\section*{Acknowledgements}
We are grateful to Stefan Theisen for correspondence and discussions on the methods and results in \cite{Schwimmer:2010za}. We thank Chris Beem, Zohar Komargodski, Finn Larsen, Jim Liu, and Balt van Rees for helpful discussions on the physics presented here. Most of this work was done while NB was a postdoc at the Simons Center for Geometry and Physics and he would like to thank this institution for its support and great working atmosphere. The work of NB is supported by Perimeter Institute for Theoretical Physics. Research at Perimeter Institute is supported by the Government of Canada through Industry Canada and by the Province of Ontario through the Ministry of Research and Innovation. HE is supported by NSF CAREER Grant PHY-0953232 and by a Cottrell Scholar Award from the Research Corporation for Science Advancement. Both HE and TMO are supported in part by the US Department of Energy under DoE Grant \#DE-SC0007859. TMO is supported by a Regents Fellowship from the University of Michigan and a National Science Foundation Graduate Research Fellowship under Grant \#F031543.

\appendix 

\section{Conformal anomaly}
\label{app:anomaly}

The conformal anomaly in four-dimensional CFTs in the presence of background metric and gauge field is well-known (see, for example, \cite{Erdmenger:1996yc} for a summary). 
The goal here is to show how this result arises from imposing the WZ consistency condition \cite{Wess:1971yu} and compatibility with the anomaly for the global $U(1)$ symmetry associated with the background gauge field.

The trace anomaly $\T$ should be a function only of the background fields $g_\mn$, $A_\m$, and their derivatives. Since the gauge symmetry is broken, it is conceivable that one could have new gauge-noninvariant contributions to the trace anomaly in addition to the standard $W^2$, $E_4$, $(F_\mn)^2$, and $\square R$ terms.\footnote{The $\square R$ ``anomaly'' is non-physical because it can be removed by a local counterterm, but we include it here for completeness. } The possible new quantities should be constructed out of the following list with various choices of the coefficients $d_i$:
\begin{align}
&d_1 \nabla_\mu(R)A^\mu + d_2\, R\, \nabla_\mu A^\mu
 + d_3 \nabla_\mu\square A^\mu  + d_4 R^{\mu\nu}\,\nabla_\mu A_\nu + d_5 R\,(A_\mu)^2 + d_6 R_{\mu\nu}A^\mu A^\nu
 \nonumber\\
&  + d_7 \nabla_\mu(A^\mu) \nabla_\nu(A^\nu)
+ d_8 \nabla_\mu(A^\nu) \nabla_\nu(A^\mu) + d_{9} \nabla_\mu(A_\nu) \nabla^\mu(A^\nu) + d_{10} A_\mu \nabla_\nu\nabla^\nu A^\mu + d_{11} A^\nu \nabla_\mu \nabla_\nu A^\mu
\nonumber\\
& + d_{12} (A_\mu)^2\, \nabla_\nu A^\nu + d_{13} A^\mu \,A^\nu\, \nabla_\mu A_\nu + d_{14}\, (A_\mu)^4 \;.
\label{anoms}
\end{align}
We will find, however, that none of these possibilities are allowed in the trace anomaly.

\vspace{6mm}
\compactsubsection{WZ consistency conditions}
The full action $S$ should satisfy the Wess-Zumino consistency conditions \cite{Wess:1971yu} (see also \cite{Cappelli:1988vw} for further discussion). In particular, since the Weyl variation of $S$ is the 
trace anomaly, the WZ conditions amount to the requirement
\begin{equation}
\int d^4x\, \Big(\s_2 \delta_{\s_1} - \s_1\delta_{\s_2} \Big) \sqrt{-g}\, \langle T_\mu{}^\mu\rangle = 0 \;.
\end{equation}
The usual anomalies, $W^2$, $E_4$, $(F_\mn)^2$, and $\square R$, satisfy that constraint, but it remains to check whether any combination of the terms in $\eqref{anoms}$ might also work. In fact, one can verify that each of the following independently satisfies the constraint:
\begin{equation}\label{Kdef}
\begin{split}
K_1 &=  \nabla_\mu \Big(3 R^{\mu\nu}\,A_\nu - R\, A^\mu + 3 \square A^\mu \Big)\;,\\
K_2 &=  \nabla_\mu \Big( A^\nu\, \nabla_\nu A^\mu \Big)\;,\\
K_3 &=  \nabla_\mu \Big( A^\mu\, \nabla_\nu A^\nu \Big)\;,\\
K_4 &=  A_\mu\, \nabla_\nu(F^{\mu\nu})\;,\\
K_5 &=  \nabla_\mu \Big(A^\mu\,(A_\nu)^2\Big)\;,\\
K_6 &= (A_\m)^4\;.\\
\end{split}
\end{equation}
Therefore based on the WZ consistency conditions alone, the trace anomaly can take the form
\begin{align}
c W^2 - a E_4 + b' \square R + \kappa_0 (F_{\mu\nu})^2 + \kappa_1 K_1 + \kappa_2 K_2  + \kappa_3 K_3 + \kappa_4 K_4 +\kappa_5 K_5 + \kappa_6 K_6 \;,
\label{generalAnomaly}
\end{align}
where the first four terms are the standard conformal anomalies in the presence of a background gauge field and curved background for a theory with central charges $c$ and $a$ \cite{Erdmenger:1996yc}. The coefficient of $(F_{\mn})^2$ is generally an independent physical quantity, although for $\mathcal{N}=1$ theories it is fixed in terms of $c$ and $a$.

\vspace{6mm}
\compactsubsection{Constraints on $\T$ from the gauge anomaly}
Just as the Weyl anomaly does not depend on either $\tau$ or $\beta$, the gauge anomaly \eqref{gaugevary} should also be a function of just the background fields. Thus there cannot be gauge dependent fields in 	\reef{generalAnomaly}; under a gauge variation those terms generate $\tau$-dependent contributions to the gauge anomaly. To illustrate this point, let us consider an example. Suppose $\kappa_6\neq 0$, so $\T$ includes an $(A)^4$ anomaly. Since $\sqrt{-g}(A)^4$ is Weyl invariant, the action whose variation produces this anomaly is simply 
\begin{equation}
S_{\WZ,A^4} = \kappa_6\int d^4x\,\sqrt{-g}\,\tau\,(A)^4\;.
\end{equation}
Now consider a gauge variation of this action, which should produce the gauge anomaly as in \eqref{gaugevary}
\begin{equation}
\delta_\alpha S_{\WZ,A^4} \sim \kappa_6\int d^4x\,\sqrt{-g}\,\tau\,(A)^3\,\D\alpha \;,
\end{equation}
which is $\tau$-dependent. The other new quantities have similar issues; in fact, no linear combination of $K_1,\ldots,K_6$ in \eqref{Kdef} is gauge invariant. This forces us to set $\kappa_1=\kappa_2=\ldots=\kappa_6 = 0$ so that the trace anomaly is gauge invariant.

Since none of the new possibilities can contribute, we find that the trace anomaly for any $\mathcal{N}=1$ superconformal theory is
\begin{equation}
\langle T_\mu{}^\mu \rangle = 
c W^2 - a E_4 + b' \square R - 6\,c\, (F_{\mu\nu})^2 \;,
\end{equation}
where the coefficient $\kappa_0=-6c$ of the last term is fixed by supersymmetry as in \cite{Schwimmer:2010za,Anselmi:1997am,Cassani:2013dba} (though with different normalization for the gauge field).



\begin{thebibliography}{9}
 
\bibitem{Schwimmer:2010za} 
  A.~Schwimmer and S.~Theisen,
  ``Spontaneous Breaking of Conformal Invariance and Trace Anomaly Matching,''
  Nucl.\ Phys.\ B {\bf 847}, 590 (2011)
  [arXiv:1011.0696 [hep-th]].
 
\bibitem{Komargodski:2011vj} 
  Z.~Komargodski and A.~Schwimmer,
  ``On Renormalization Group Flows in Four Dimensions,''
  JHEP {\bf 1112}, 099 (2011)
  [arXiv:1107.3987 [hep-th]].

\bibitem{Komargodski:2011xv} 
  Z.~Komargodski,
  ``The Constraints of Conformal Symmetry on RG Flows,''
  JHEP {\bf 1207}, 069 (2012)
  [arXiv:1112.4538 [hep-th]].

\bibitem{Luty:2012ww} 
  M.~A.~Luty, J.~Polchinski and R.~Rattazzi,
  ``The $a$-theorem and the Asymptotics of 4D Quantum Field Theory,''
  JHEP {\bf 1301}, 152 (2013)
  [arXiv:1204.5221 [hep-th]].
  
\bibitem{Dymarsky:2013pqa} 
  A.~Dymarsky, Z.~Komargodski, A.~Schwimmer and S.~Theisen,
  ``On Scale and Conformal Invariance in Four Dimensions,''
  arXiv:1309.2921 [hep-th].
  
\bibitem{Farnsworth:2013osa} 
  K.~Farnsworth, M.~A.~Luty and V.~Prelipina,
  ``Scale Invariance plus Unitarity Implies Conformal Invariance in Four Dimensions,''
  arXiv:1309.4095 [hep-th].

\bibitem{Anselmi:1997am} 
  D.~Anselmi, D.~Z.~Freedman, M.~T.~Grisaru and A.~A.~Johansen,
  ``Nonperturbative formulas for central functions of supersymmetric gauge theories,''
  Nucl.\ Phys.\ B {\bf 526}, 543 (1998)
  [hep-th/9708042].

\bibitem{Intriligator:2003jj} 
  K.~A.~Intriligator and B.~Wecht,
  ``The Exact superconformal R symmetry maximizes a,''
  Nucl.\ Phys.\ B {\bf 667}, 183 (2003)
  [hep-th/0304128].

\bibitem{Cassani:2013dba} 
  D.~Cassani and D.~Martelli,
  ``Supersymmetry on curved spaces and superconformal anomalies,''
  arXiv:1307.6567 [hep-th].
 
\bibitem{Elvang:2010jv}
  H.~Elvang, D.~Z.~Freedman and M.~Kiermaier,
  ``A simple approach to counterterms in N=8 supergravity,''
  JHEP {\bf 1011}, 016 (2010)
  [arXiv:1003.5018 [hep-th]].
 
\bibitem{Grisaru:1976vm} 
  M.~T.~Grisaru, H.~N.~Pendleton and P.~van Nieuwenhuizen,
  ``Supergravity and the S Matrix,''
  Phys.\ Rev.\ D {\bf 15}, 996 (1977).

\bibitem{Grisaru:1977px} 
  M.~T.~Grisaru and H.~N.~Pendleton,
  ``Some Properties of Scattering Amplitudes in Supersymmetric Theories,''
  Nucl.\ Phys.\ B {\bf 124}, 81 (1977).
 
\bibitem{Elvang:2013cua} 
  H.~Elvang and Y.-t.~Huang,
  ``Scattering Amplitudes,''
  arXiv:1308.1697 [hep-th].
 
\bibitem{Dixon:2013uaa} 
  L.~J.~Dixon,
  ``A brief introduction to modern amplitude methods,''
  arXiv:1310.5353 [hep-ph].

\bibitem{Wess:1971yu} 
  J.~Wess and B.~Zumino,
  ``Consequences of anomalous Ward identities,''
  Phys.\ Lett.\ B {\bf 37}, 95 (1971).
  
\bibitem{Elvang:2012st} 
  H.~Elvang, D.~Z.~Freedman, L.~-Y.~Hung, M.~Kiermaier, R.~C.~Myers and S.~Theisen,
  ``On renormalization group flows and the a-theorem in 6d,''
  arXiv:1205.3994 [hep-th].
  
\bibitem{Elvang:2012yc} 
  H.~Elvang and T.~M.~Olson,
  ``RG flows in d dimensions, the dilaton effective action, and the a-theorem,''
  JHEP {\bf 1303}, 034 (2013)
  [arXiv:1209.3424 [hep-th]].
 
\bibitem{Baume:2013ika} 
  F.~Baume and B.~Keren-Zur,
  ``The dilaton Wess-Zumino action in higher dimensions,''
  arXiv:1307.0484 [hep-th].

\bibitem{Erdmenger:1996yc} 
  J.~Erdmenger and H.~Osborn,
  ``Conserved currents and the energy momentum tensor in conformally invariant theories for general dimensions,''
  Nucl.\ Phys.\ B {\bf 483}, 431 (1997)
  [hep-th/9605009].

\bibitem{Cappelli:1988vw} 
  A.~Cappelli and A.~Coste,
  ``On The Stress Tensor Of Conformal Field Theories In Higher Dimensions,''
  Nucl.\ Phys.\ B {\bf 314}, 707 (1989).
    
\end{thebibliography}
\end{document}